\documentclass[preprint,12pt]{elsarticle}

\usepackage{amssymb}
\usepackage{amsmath}

\usepackage{epsfig}
\usepackage{url}
\usepackage{cite}
\usepackage{fancybox}
\usepackage{multirow}
\usepackage{makecell}
\usepackage[table,xcdraw]{xcolor}
\usepackage[utf8]{inputenc}
\usepackage[T1]{fontenc}

\journal{Nuclear Physics B}

\begin{document}

\begin{frontmatter}



\author{Larissa Barbosa\corref{cor1}\fnref{label1}}
 \ead{larissa.leoncio@ufba.br}
 \cortext[cor1]{Corresponding author}
 \affiliation[label1]{organization={Federal University of Bahia},
            state={Bahia},
            country={ Brazil}}

\author[label2,label3]{Sávio Freire}
 \ead{savio.freire@ifce.edu.br}
 \affiliation[label2]{organization={Federal Institute of Ceará},
             state={Ceará},
             country={Brazil}}
 \affiliation[label3]{organization={State University of Ceará},
             state={Ceará},
             country={Brazil}}
             
 \author[label4]{Marcos Kalinowski}
\ead{kalinowski@inf.puc-rio.br}
 \affiliation[label4]{organization={Pontifical Catholic University of Rio de Janeiro},
            state={Rio de Janeiro},
            country={ Brazil}}

 \author[label5]{Zadia Codabux}
\ead{zadiacodabux@ieee.org}
 \affiliation[label5]{organization={University of Saskatchewan},
            state={Saskatoon},
            country={Canada}}

 \author[label6]{Rodrigo Spínola}
\ead{spinolaro@vcu.edu}
 \affiliation[label6]{organization={Virginia Commonwealth University},
            state={Richmond},
            country={United States}}
            
 \author[label1]{Manoel Mendonça}
\ead{manoel.mendonca@ufba.br}

\author[label1]{Rita S. P. Maciel}
 \ead{rita.suzana@ufba.br}

\title{Understanding Undesirable Attributes of Requirements Engineers: Insights from Practitioners} 


\author{} 


\begin{abstract}
\textbf{Context.}  The characteristics of software professionals have been widely investigated in the literature. However, limited attention has been given to undesirable attributes in Requirements Engineering, despite the strong dependence of this activity on stakeholder interaction and collaboration.
\textbf{Objectives.} This study investigates the undesirable attributes of requirements engineers' hat may hinder collaboration and project success. 
\textbf{Method.} We surveyed software practitioners to identify these attributes and conducted interviews to gather supporting evidence. 
\textbf{Results.} Seventeen undesirable attributes were identified, grouped into four categories (communication issues, lack of domain knowledge, personality, and lack of technical knowledge), and organized into conceptual maps. 
\textbf{Conclusion.} The maps help requirements engineers reflect on and improve their professional practice by recognizing traits that may hinder collaboration and project outcomes.
\end{abstract}

\begin{keyword}
requirements engineer \sep undesirable attributes \sep survey \sep interview
\end{keyword}

\end{frontmatter}

\section{Introduction}
Requirements engineers bridge human needs and technical solutions, translating stakeholder expectations into actionable requirements. The activity performed by the requirements engineer depends heavily on the interaction between stakeholders, team members, and undesirable behaviors or attitudes can compromise collaboration and, ultimately, project outcomes.
We have previously investigated issues related to software requirements [1] and surveyed practitioners to understand the attributes of an effective requirements engineer [2]. However, little is known about undesirable attributes. 
Understanding them is essential to prevent harmful behaviors and enhance the effectiveness of RE and provide actionable insights that help organizations improve recruitment, training, and teamwork in RE. Therefore, we seek to answer: ``\textit{What are the less desirable attributes of a requirements engineer?}''

\section{Research Method}
As we did not have a previous list of attributes, we first designed a survey to identify desirable and undesirable attributes of requirements engineers. After, we invited eighteen software practitioners from our Brazilian industrial partners, but only 11 of them were available for interviews and for answering the follow up question. This approach was adopted from Kalliamvakou et al. [3] and Dias et al. [4].
Participants had between one and thirty years of experience (average = 10 years) and had worked for an average of six companies (range: one to twelve). They represented organizations of different sizes: small (up to 50 employees; n = 2), medium (51–1,000 employees; n = 6), and large (more than 1,000 employees; n = 10). The sample included ten software engineers and eight project managers.
We decided to invite requirements engineers and project managers to capture the perception of who performs RE activities and who manages them. Each practitioner listed up to five attributes they considered essential for performing RE activities and up to five they considered detrimental. Two researchers independently analyzed the attributes to identify similarities, with divergences resolved by a third researcher. One researcher grouped related attributes into categories, which were reviewed by another researcher.
Subsequently, we remotely interviewed eleven of the surveyed practitioners to clarify each attribute’s definition. Participants also explained why the attribute was considered positive or negative and how it could be recognized or avoided. The interviews lasted about 30 minutes, and were recorded with participants’ consent. After transcribing the interviews, the same researchers applied open coding [5] of the transcripts to capture the main ideas in each response. Each of them, individually, coded half of the transcripts and then revised each other. Afterwards, they met with the last author to resolve the divergences, resolving disagreements with a third researcher. In the end, we produced a consolidated set of codes representing the key concepts of each attribute.

While our previous work [2] reported the desirable attributes of requirements engineers, this current study focuses on identifying and organizing the undesirable attributes.

\section{Results}
\begin{figure*}
\centering
\makebox[\textwidth][c]{%
\includegraphics[scale=0.1]{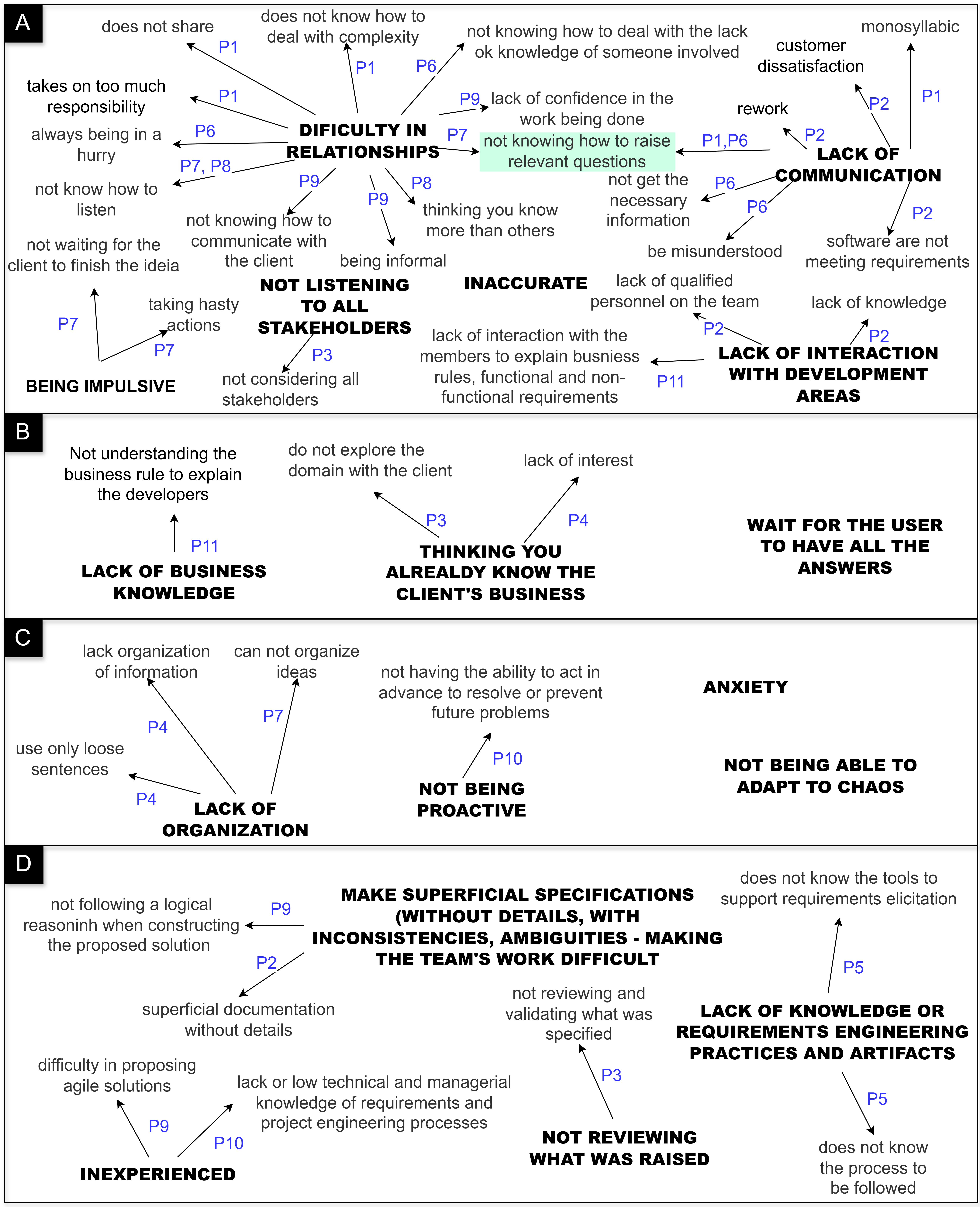}
}
\caption{Map for (A) Communication Issues, (B) Lack of Domain knowledge, (C) Personality, and (D) Lack of Technical Knowledge  categories. In the maps, attributes are represented in bold upper-case text, whereas the concepts are represented as smaller, lowercase text. An attribute has a set of concepts that describe it. An arrow relates an attribute and its concepts. It is worth noting that not all 17 attributes have their definitions, as some were not addressed during the interviews.}
\label{FIG:framework}
\end{figure*}

Fig.~\ref{FIG:framework} presents a set of maps that organize the undesirable attributes. Each map groups a set of attributes within a specific category, following the framework proposed by Dias et al. [4].

Seventeen undesirable attributes emerged\footnote{The complete list of undesirable attributes with their corresponding number of mentions and quotes is available at \url{https://zenodo.org/records/20089943}} with relationship and communication difficulties beeing as the most recurrent. These were often interrelated, as poor communication could exacerbate relational challenges. The following attributes were the most cited: (i) \textit{difficulty in relationships} (18\%) means the challenges that affect relationships with team memebers and stakeholders; (ii) \textit{lack of communication} (15\%) refers to the absence or deficiency in the exchange of information, ideas, and feedback between individuals or teams, resulting in misunderstandings, lack of coordination, and inefficiencies, and (iii) \textit{lack of business knowledge} (8\%) is related to the lack of adequate understanding of the context, objectives, and needs of the business for which the software is being developed. Navigating on the map, we can see that \textit{not knowing how to raise relevant questions} is part of the definition of the following attributes \textit{difficulty in relationships}, \textit{lack of communication}, indicating that these attributes may be complements. 

In addition, the attributes were grouped into four categories: a) \textbf{Communication Issues}, expressing issues on how requirements engineers communicate (e.g., \textit{difficulty in relationships}), b) \textbf{Lack of Domain Knowledge}, grouping attributes (e.g., \textit{lack of business knowledge}) to reduce the domain and business knowledge; c) \textbf{Lack of Technical knowledge}, revealing lack of technical skills that are used in RE activities (e.g., \textit{lack of knowledge of RE practices and artifacts}); and d) \textbf{Personality}, with attributes related to express how requirements engineers think, feel, and behave during RE activities and interactions with stakeholders (e.g., \textit{impressive}).

Overall, the conceptual maps highlight how technical deficiencies and interpersonal issues intertwine, suggesting that undesirable traits in RE are not merely individual flaws, but systemic challenges in team dynamics and communication.

\section{Conclusion}
Our findings align with previous studies that highlight the importance of human factors in RE. Being considered a good software professional requires a combination of technical and non-technical skills, commonly referred to as technical skills and (soft)interpersonal skills, respectively. These competencies include teamwork, problem-solving, communication, and leadership [6, 7]. In particular, empirical studies have indicated that interpersonal skills, such as communication, analytical thinking, and teamwork, are often more in demand than technical skills in job descriptions in the field [8]. In this sense, this study reinforces this evidence from a complementary perspective, empirically showing that many undesirable attributes of requirements engineers are directly associated with communication and relationship problems. One point particularly caught our attention. While Niknafs and Barry [8] suggested that combining team members with and without domain knowledge can improve idea generation and team effectiveness, our findings identified lack of business knowledge as an undesirable attribute for requirements engineers.

We also compared the desirable attributes [2] with undesirable ones, showing that both are associated with communication, collaboration, organization, stakeholder interaction, and adaptability, highlighting the importance of interpersonal and organizational competencies in requirements engineering. However, the findings also indicate that undesirable attributes are not simply the negation of desirable ones. While some contrasts exist, such as communication skills versus lack of communication, several undesirable attributes involve restrictive behaviors, such as resistance to change and poor interpersonal relationships, whereas desirable attributes are associated with proactive behaviors, including initiative, negotiation, and problem-solving. Overall, the results suggest that desirable and undesirable attributes represent different dimensions of professional behavior rather than direct opposites.

However, we acknowledge that our sample comprises 18 participants from the Brazilian software industry may introduce contextual bias. Shared industry practices, as well as geographical and cultural factors, may influence participants’ perceptions. To mitigate this threat, we included professionals with diverse experience levels, company sizes, and career trajectories. Furthermore, Brazil’s economic and cultural diversity may partially reduce regional bias. We discuss other threats in the complementary material (\url{https://zenodo.org/records/20089943}).

\end{document}